# AI-driven random walk simulations of viscophoresis and visco-diffusiophoretic particle trapping

K. Mathwig

*imec within OnePlanet Research Center, Bronland 10, 6708 WH Wageningen, the Netherlands*

klaus.mathwig@imec.nl

24 January 2025

Viscophoresis refers to the transport of suspended nanoparticles driven by a steep viscosity gradient. This work investigates this new transport effect using a random walk simulation. By modelling position-dependent Brownian motion, viscophoresis, and diffusiophoresis in a one-dimensional geometry, the simulation yields results that align well with experimental data, demonstrating viscophoresis as a new phoretic transport mechanism. Additionally, the simulation predicts the efficient separation of nanoparticles based on size, suggesting potential applications for sorting in microfluidic systems. The Python script for the simulation was generated using ChatGPT o1, significantly accelerating model development and providing accurate physical insights and efficient equations. However, caution is advised, as ChatGPT may generate non-physical results; iterative testing and validation is important.

Nonequilibrium systems can exhibit multiplicative noise, where the magnitude of random fluctuations scales with a signal and depends on the state of the system. Interpreting this type of noise presents the *Itô-Stratonovich dilemma*, a challenge that occurs in fields as diverse as finance and electrical engineering. This dilemma was a major topic in statistical physics during the 1970s and 1980s,[1] and it remains a subject of active research.[2–4]

A straightforward example is diffusion of a suspended particle undergoing Brownian motion in a gradient of diffusivity. The length of each step of the particle's random walk depends on the local diffusivity, raising the central question of the Itô-Stratonovich dilemma: where should this diffusivity be evaluated—at the particle's position at the beginning of the step or along the way? This dilemma arises because the underlying Langevin equation can be interpreted in multiple ways, making it impossible to resolve this nonequilibrium diffusion from first principles alone.

Recently, we investigated nanoparticle diffusion in such conditions by experimentally generating a steep diffusivity gradient in a microfluidic channel, created through a viscosity gradient in the suspension medium. Our results[5] reveal strong *viscophoresis*, where particles drifted along the diffusivity gradient with velocities exceeding by far those predicted by conventional interpretations of Itô-Stratonovich diffusion.

In this paper, I use a one-dimensional random walk simulation to model both diffusion-gradient-driven drift and particle trapping. The simulation shows strong agreement with experimental data and offers predictions for new experimental parameters.

The Python script used for this simulation was generated entirely using ChatGPT o1-preview, which facilitated rapid model development. While ChatGPT o1 provided useful suggestions regarding the physics reasoning, careful validation of its output is essential.

The concept of viscophoresis is shown in Figure 1a. A nanoparticle within a viscosity gradient in a microfluidic channel experiences a linear gradient in its diffusion coefficient, which varies from one end to the channel to the other. Consequently, steps of the random walk are longer on the right side (where diffusivity is higher) and shorter of the left side (where diffusivity is lower). For a one-dimensional walk, the length of each diffusive step is given by $\Delta x = \sqrt{2D\Delta t}$, where $D = D(x)$ is the position-dependent diffusion coefficient. Tackling the Itô-Stratonovich dilemma involves choosing the appropriate form for $D(x)$, with the three typical choices (see Fig. 2):[2]

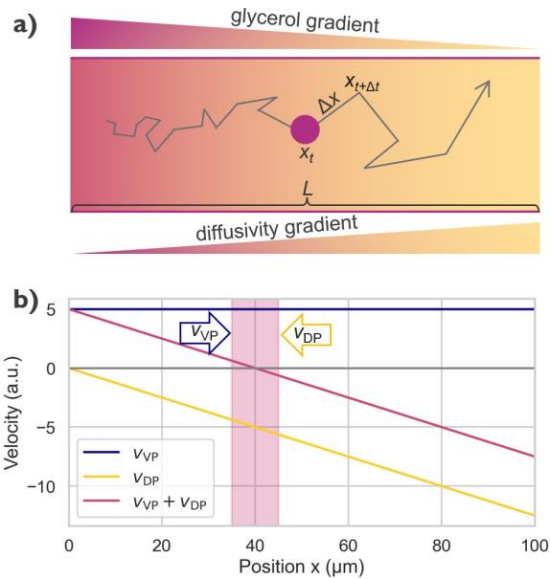

FIG. 1. Schematic of viscophoretic drift and particle trapping. a) A stable gradient of glycerol in an aqueous solution in a microfluidic channel (length $L$) generates a steep viscosity gradient, resulting in a linear gradient in diffusivity $D$. A nanoparticle undergoing Brownian motion takes longer steps in the direction of higher diffusivity. b) The asymmetry in step lengths leads to a constant viscophoretic drift $v_{\mathrm{VP}}$ (blue curve). Simultaneously, the glycerol gradient induces a diffusiophoretic drift $v_{\mathrm{DP}}$ (yellow curve) in the opposite direction. As $v_{\mathrm{DP}}$ increases linearly, the net velocity (purple) reaches zero around 40 μm, and particles start to accumulate at this location (shaded area).

$$\Delta x = \sqrt{2D(x_t)\Delta t} \quad \text{(Itô)},$$
$$\Delta x = \sqrt{2D\left(\frac{x_t + x_{t+\Delta t}}{2}\right)\Delta t} \quad \text{(Stratonovich)}, \quad (1)$$
$$\Delta x = \sqrt{2D(x_{t+\Delta t})\Delta t} \quad \text{(isothermal)}.$$

AI-driven random walk simulations of viscophoresis and visco-diffusiophoretic particle trapping

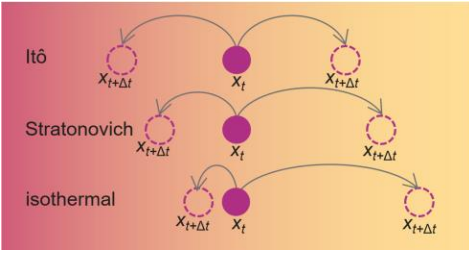

FIG 2. Schematic of different step length of nanoparticles undergoing Brownian motion in a gradient of viscosity and diffusivity. Particles are undergoing a random walk with stochastic steps according to Itô, Stratonovich, and isothermal diffusion. Figure adapted from arXiv:1807.09106 [physics.flu-dyn][2].

In other words, the step length $\Delta x$ is determined by the diffusion coefficient at the beginning (Itô), middle (Stratonovich), or end (isothermal) of the step. These choices directly impact the magnitude of a diffusive (or viscophoretic) drift velocity:

$$v_{\text{VP}} = \langle \dot{x} \rangle = \alpha \frac{dD(x)}{dx}, \quad (2)$$

where $\alpha = \{0, ½, 1\}$ corresponds to the Itô, Stratonovich, and isothermal interpretations, respectively. As shown in Figure 2, there is no drift in the Itô case, but considerable drift can arise in the isothermal case, given by $v_{\text{VP}} = (D_{x=0} - D_{x=L})/L$. ($L$: gradient or channel length).

Experimentally, viscophoretic drift cannot be observed in complete isolation from other effects. In our experiment, the diffusivity/viscosity gradient was created by varying the concentration of glycerol in an aqueous solution, which also induced *diffusio*phoretic[6,7] drift of the suspended nanoparticles. I.e., the particles are driven by the steep gradient of glycerol as glycerol molecules diffuse around them particles. Unlike viscophoresis, diffusiophoretic velocity is not constant but increases linearly as the glycerol concentration decreases. As illustrated in Figure 1b, visco- and diffusiophoretic drifts act in opposite directions.[5] As a result, the net drift velocity reaches zero at a specific point within the microchannel, where particles accumulate. This is because particles are transported from both sides—driven by dominant viscophoresis on the left side and dominant diffusiophoresis on the right.

We observed such trapping for nanoparticles ranging from 10 nm to 100 nm in diameter.[5] Using particle image velocimetry, we mapped both diffusivities and velocities,[8] finding very fast viscophoretic drift with values up to up to $v_{\text{VP}} = 5.8 \, \mu m/s = 34 \cdot dD(x)/dx$ (i.e., $\alpha = 34$) for particles with a 28 nm diameter.

Such as strong viscophoretic drift velocity of $\alpha > 1$ has not previously been observed. Thus, no analytical or numerical methods exist to describe the observed effect.

The stochastic simulation adds value in several ways:

1. It allows comparison of experimental and simulated particle distributions, with agreement between the two supporting the experimental interpretations.
2. It enables predictions for particle behaviour under new experimental conditions.
3. Specifically, the simulation helps predict particle *separation* behaviour.

A Python script for a random walk is provided in Appendix 2 (see also https://dx.doi.org/10.6084/m9.figshare.27200823). Particles are initially distributed randomly along a channel of length $L$ and begin a one-dimensional random walk with a step length $\Delta x = \sqrt{2D(x)\Delta t}$. Particles stepping out and exiting the channel are removed from the simulation. Meanwhile, coupling to reservoirs is considered: particles enter the channel at both edges at a specified frequency and begin their random walk. In addition to the random walk, particles experience position-dependent diffusiophoretic drift, $v_{\text{DP}}$, which varies linearly with their current position. Superimposed on this is a constant viscophoretic drift $v_{\text{VP}} = \alpha \, (D_{x=0} - D_{x=L})/L$.

The code of the simulation was not written manually but generated entirely by prompting ChatGPT o1-preview (OpenAI, USA), a generative pre-trained transformer [9] (GPT) large language model (LLM) released just four weeks prior to writing this paper. Thus, I share my experience to illustrate the utility of this new tool. Using ChatGPT significantly accelerated the process, allowing the script to be generated in less than one working day. I had not any previous experience with Python or using large language models. However previous experience in modelling Brownian random walks[10,11] in MATLAB was helpful for understanding the generated script.

ChatGPT was used as a "co-intelligence"[12] tool in an iterative process. I started by generating a simple random walk, then gradually added boundary conditions and drift terms step by step. At each stage, I checked and debugged the code to ensure that the physics was implemented correctly. It is important to note that there is always the risk of ChatGPT "hallucinations", where the model confidently generates incorrect or misleading outputs. Therefore, careful oversight is essential to ensure accuracy and the correct implementation of drift and diffusion.

It is claimed that ChatGPT o1-preview can "… reason through complex tasks and solve harder problems than previous models in science, coding, and math."[13] While the underlying mechanisms of this "reasoning"[14] are proprietary and not fully disclosed, the model provided several useful improvements to the random walk simulation. For example, where I would have used a fixed step length $\Delta x = \sqrt{2D\Delta t}$, ChatGPT suggested a more realistic approach by drawing a random sample from a Gaussian distribution with standard deviation $\sqrt{2D\Delta t}$.

More impressively, when ChatGPT was prompted to calculate viscophoretic drift by evaluating step length via anticipated positions (sampling $D(x)$ at the position at the end of the step, and then determining step length by using an updated $D$, see Eq. (1)), it recognized this as an attempt at Itô-Stratonovich diffusion and suggested a more elegant implementation using Eq. (2) instead. The response and "reasoning" are quoted in Appendix 1.

Results of the random walk including are shown in Figure 3. Figure 3a shows time traces of 20 particles. Initially positioned randomly within a 100 μm-long channel, the particles are being focused into a narrow region due to dominant viscophoresis in the range $x = 0…25$ μm and diffusiophoretic drift in the range $x = 25…100$ μm. Figure 3b shows histograms of the resulting particle distribution over time. The total number of particles increases steadily due to the generation of 100 particles per second at both edges of the simulation (with 1000 particles initially within the channel).

The simulation parameters were chosen based on the following considerations:

- Diffusion coefficients and gradients: The diffusion coefficients were calculated using the Stokes–Einstein–Sutherland equation $D = k_B T/(6\pi r \eta)$ ($k_B$: Boltzmann constant, $T$: temperature, $r$: particle radius, $\eta$: dynamic

AI-driven random walk simulations of viscophoresis and visco-diffusiophoretic particle trapping

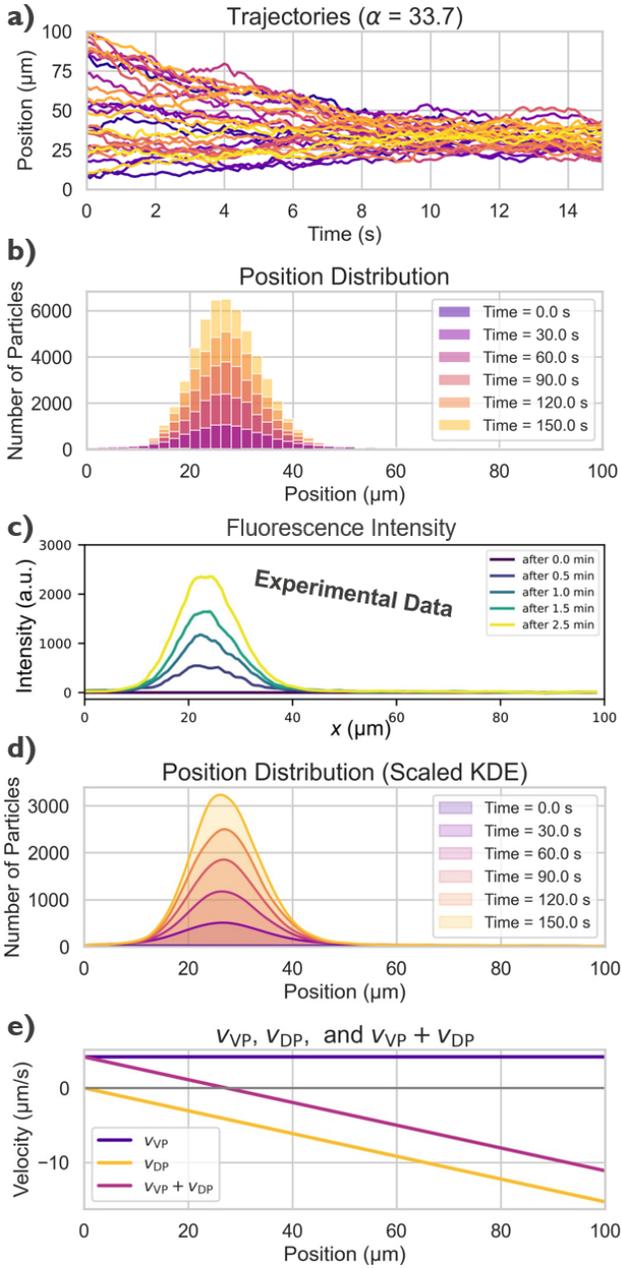

FIG. 3. a) Random walk trajectories of 20 particles starting from random positions. b) Histogram showing the particle distribution at different time points. c) Experimental results for 28 nm particles under the same conditions; fluorescence intensity of particles accumulating in a 100 μm-long microchannel is shown. d) Same data as in panel b) plotted as curves (scaled KDE) for better comparison with experimental data in panel c). e) Corresponding velocities $v_{VP}$, $v_{VP}$ and net velocity. Panel c) is taken from arXiv:2212.11503 [physics.flu-dyn][5] (CC BY).

- Viscophoretic drift: Experimentally determined values of the α parameter were used for the viscophoretic drift velocity as no theoretical or analytical predictions exist yet for drift faster than α = 1. For 28-nm particles in a microchannel of length $L$ = 100 μm with a diffusivity gradient of $\Delta D/L$ = 0.12 μm/s, the viscophoretic drift velocity is calculated to be to $v_{VP} = 34 \cdot \Delta D/L = 4.1$ μm/s (α = 34; see blue line in Figure 3e).

- Diffusiophoretic drift: Diffusiophoretic drift has primarily been studied in salt solution gradient as an electrokinetic effect.[6] However, nonelectrolyte diffusiophoresis remains less understood, with only a few experimental studies available.[7,16] The magnitude of $v_{DP}$ cannot readily be calculated from first principle, so we rely on experimentally determined values. The drift velocity magnitude increases nearly from $v_{DP}(x=0) = 0$ to the maximum observed velocities at $v_{DP}(x=L)$. Experimentally, $v_{DP}(x = L = 100$ μm$)$ = -15 μm/s for both 28 nm and 110 nm particles.

Experimental results of fluorescent intensities, corresponding to the concentration 28-nm particles, are shown in Figure 3c. For better visual comparison, the simulated values are plotted as continuous densities using a kernel density estimate (KDE) in Figure 3d. All parameters—trapping position, distribution shape, and time evolution—show good agreement between experimental data and the one-dimensional random walk simulation.

Given that both viscophoretic and diffusiophoretic drifts were simulated using experimentally derived parameters, it is unsurprising that the particle trapping position at $x$ = 25 μm matches well. Nonetheless, the diffusive broadening of particle accumulations cannot be easily predicted using simple analytical estimates (see Figure 3e). Thus, the random walk simulation strongly supports the interpretation of experimental experiments as interplay of viscophoresis and diffusiophoresis.

The random walk simulation demonstrates predictive power in determining optimal parameters for manipulating nanoparticle transport. For instance, viscophoresis and diffusiophoresis can be leveraged to *separate* particles based on their size and hydrodynamic radius. This concept is illustrated in Figure 3, which shows the distribution of 28-nm and 110-nm particles after 2 min of trapping, using experimentally determined $v_{VP}$ values for both particle sizes. While larger 110-nms particle have a higher α = 46, their viscophoretic drift velocity is slower due to their lower diffusivity. As a result, they accumulate on the far-left side of the microchannel and exhibit a narrower distribution because of the slower diffusion and shorter step lengths, respectively.

In Figures 4a-c, the length of the microchannel is varied while keeping other parameters constant. A shorter 50-μm long channel (Figure 4b) with steeper gradients leads to better separation of the two particle sizes, achieving near baseline separation. However, when the channel length is further reduced to $L$ = 35 μm, the higher diffusivity and faster viscophoresis of the 28-nm particles causes them to be flushed out of the channel, preventing accumulation.

Overall, trapping and separation are limited to the regime of micrometer-scale channels and nanometer-scale particles. For small molecules and mono-atomic ions,[2] viscophoresis will dominate over diffusiophoresis due to their high diffusivities. Similarly, shrinking the channel size to the nanometer range (e.g., using a nanopore[17]) would result in extremely fast viscophoretic drift velocities, preventing any trapping from occurring.

viscosity). For example, an aqueous solution with glycerol concentration from 50 wt% to 0 wt% ranges in viscosity from 6 mPa·s to 1 mPa·s.[15] Particles with a 28-nm diameter suspended in this solution gradient experience diffusivities ranging from $2.6 \cdot 10^{-12}$ m²/s to $1.5 \cdot 10^{-11}$ m²/s. Calculated diffusion coefficient match well with experimental observations.[5,8]

AI-driven random walk simulations of viscophoresis and visco-diffusiophoretic particle trapping

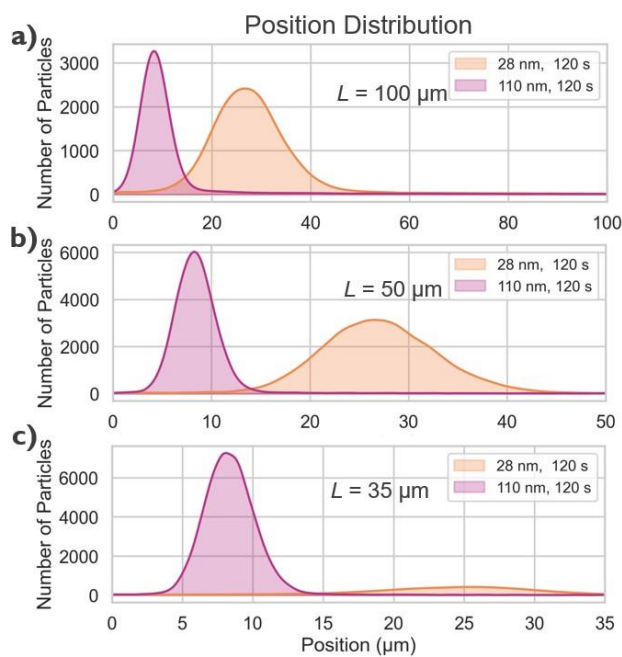

FIG. 4. Simulated particle separation based on size. Distributions of 28 nm and 110 nm diameter particles are shown after 2 minutes of drift and diffusion in a viscosity gradient ranging 6 mPa·s to 1 mPa·s. The length of the microchannels is varied from a) $L$ = 100 μm, to b) $L$ = 50 μm, and c) $L$ = 35 μm.

In this work, I used a one-dimensional random walk simulation to model Brownian motion, viscophoresis, and diffusiophoresis of nanoparticles in a microfluidics viscosity gradient. The simulation results align well with experimental data, supporting the interpretation of these experiments as a demonstration of a new phoretic transport effect, namely viscophoresis. Extending the random walk to include simultaneous drift and diffusion for two particle types with different sizes also predicts a potential application for efficient separation and sorting. Additionally, the simulation facilitates the optimization of experimental parameters for future work.

The use of ChatGPT to generate the Python script for the random walk greatly accelerated the development process. (Without this tool, this work would not have happened). The new "reasoning" functionality in ChatGPT o1 provided valuable suggestions that improved the modelling process, such as offering correct and efficient equation for Brownian step length and isothermal drift. It also provided clear and accurate explanations of the underlying physics. I encourage others to explore ChatGPT or other large language models for applications in nonlinear dynamics, statistical physics, and fluid dynamics, or also chemistry.[18]

However, at the current state of development, caution is necessary when using these tools, as hallucinations—instances where ChatGPT confidently generates incorrect or non-physical results—are a challenge. An iterative approach, building up the complexity of the script step by step while thoroughly checking and debugging at each stage, proved effective in mitigating these challenges.

## Acknowledgement

I thank Xu Zhang (F-AI & Co.) for discussions.

## Appendix 1

A prompt and response are quoted verbatim, highlighting the "reasoning" capability of ChatGPT o1-preview.

Prompt:

"in the script update method 2 to take into account that dx should depend on the direction of the random walk."

(Here, "method 2" refers to adding viscophoretic drift to the random walk. In the process building the script, so the far step length was calculated correctly, but the script did not consider that the step length is different depending on direction of the step.)

Response ChatGPT o1-preview:

Thought for 11 seconds [...]

Certainly! The request involves updating **Method 2** in the script to account for the fact that the displacement dx should depend on the **direction** of the random walk when diffusivity is position-dependent. This is important because when diffusivity varies with position, it can induce an additional drift in the particle's motion.

AI-driven random walk simulations of viscophoresis and visco-diffusiophoretic particle trapping

## Understanding the Issue

### Position-Dependent Diffusivity and Directional Bias

In systems where the diffusivity $D(x)$ depends on position $x$, the random motion of particles is not purely symmetric. There is an induced drift that depends on the gradient of the diffusivity $D'(x)$. This means particles are more likely to move towards regions of higher or lower diffusivity, depending on how $D(x)$ changes with $x$.

### Ito vs. Stratonovich Interpretations

In stochastic calculus, when dealing with stochastic differential equations (SDEs) involving multiplicative noise (i.e., position-dependent diffusivity), the interpretation of the noise term becomes crucial. The two main interpretations are:

- **Ito Interpretation**: Assumes that the noise is non-anticipating, leading to an extra drift term proportional to the derivative of the diffusivity.
- **Stratonovich Interpretation**: Preserves the usual rules of calculus, and the extra drift term does not appear explicitly.

In physical systems, the **Stratonovich interpretation** is often more appropriate because it correctly accounts for the physical behavior of particles in varying environments.

## Updating Method 2

To account for the direction of the random walk in **Method 2**, we need to:

1. **Include the Drift Term**: Add a drift term that arises due to the spatial variation of diffusivity.
2. **Adjust the Displacement Calculation**: Modify the calculation of dx to include both the stochastic term and the drift term.

## Mathematical Formulation

### Stochastic Differential Equation

The SDE for a particle moving in one dimension with position-dependent diffusivity $D(x)$ is:

$$dx = D'(x)dt + \sqrt{2D(x)}dW_t$$

- $D'(x)dt$: The drift term arising from the spatial variation of diffusivity.
- $dfd\sqrt{2D(x)}dW_t$: The stochastic term representing random fluctuations, where $dW_t$ is the Wiener process increment.

## Calculating the Drift Term

For our linear diffusivity:

$$D(x) = D_{\text{start}} + (D_{\text{end}} - D_{\text{start}})\left(\frac{x}{L}\right)$$

The derivative $D'(x)$ is constant:

$$D'(x) = \frac{dD}{dx} = \frac{D_{\text{end}} - D_{\text{start}}}{L}$$

### Direction Dependence

The drift term $D'(x)dt$ introduces a directional bias in the particle's motion, depending on whether $D(x)$ is increasing or decreasing with $x$.

[...] (Details on updated Python are omitted in this quote here, see Appendix 2 below.)

## Understanding the Effects

### Direction Dependence

- **Drift Towards Higher Diffusivity:**
  - Since $D'(x) > 0$ (diffusivity increases with position), the drift term is positive.
  - Particles tend to drift towards regions of higher diffusivity.
- **Asymmetry in Particle Motion:**
  - The inclusion of the drift term causes an asymmetry in the random walk.
  - Particles are more likely to move in the direction where diffusivity increases.

[...]

## Physical Interpretation

- **Realistic Modeling:**
  - Including the drift term aligns the simulation with the physical behavior of particles in media with spatially varying diffusivity.
- **Fokker-Planck Equation Consistency:**
  - The updated method is consistent with the Fokker-Planck equation for diffusion with position-dependent diffusivity.

# Appendix 2

AI-generated (ChatGPT o1 prompting) Python script for one-dimensional random walks and particle accumulation due to viscophoresis and diffusiophoresis. Only The random walk generation is shown below but no the script for plotting. For the entire script used to plot Figures 3 and 4, see https://dx.doi.org/10.6084/m9.figshare.27200823.

```
import numpy as np
import matplotlib.pyplot as plt

# Parameters
L = 100.0                  # Range length in micrometers (starting from 0)
D_start = 7e-12            # Diffusivity at the beginning of the range (m²/s)
D_end = 17e-12             # Diffusivity at the end of the range (m²/s)
alpha = 38                 # Alpha parameter (0 for Stratonovich, 0.5 for Ito)
v_start = 0e-6             # Diffusiophoretic drift at the beginning of the range (m/s)
v_end = -15e-6             # Diffusiophoretic drift at the end of the range (m/s)
N_particles_initial = 2000 # Initial number of particles
```

```python
# Particle generation parameters
particles_per_edge_left = 30      # Particles generated at the left edge per generation event
generation_interval_left = 1      # Time steps between generations at the left edge

particles_per_edge_right = 30     # Particles generated at the right edge per generation event
generation_interval_right = 1     # Time steps between generations at the right edge

dt = 0.1                          # Time step in seconds
T_total = 180                     # Total simulation time in seconds
num_displayed_traces = 30         # Number of particle traces to display

# Plotting parameters for middle and bottom diagrams
first_curve_time = 0              # Time of the first curve
delta_t_curves = 30               # Time difference between curves
num_curves = 6                    # Number of curves to plot

# Convert micrometers to meters for calculations
L_meters = L * 1e-6

# Number of time steps
num_steps = int(T_total / dt)

# Initialize positions and particle IDs
positions = np.random.uniform(0, L_meters, N_particles_initial)
particle_ids = np.arange(N_particles_initial)

# Initialize a counter for assigning unique IDs to new particles
next_particle_id = N_particles_initial

# Select particles to track
tracked_particle_ids = np.random.choice(particle_ids, size=num_displayed_traces, replace=False)
trajectories = {pid: [] for pid in tracked_particle_ids}

# Lists to store data over time
positions_over_time = []
time_points = []

# Calculate the constant diffusivity gradient D'(x)
D_gradient = (D_end - D_start) / L_meters

# Calculate the constant diffusiophoretic drift gradient v'(x)
v_gradient = (v_end - v_start) / L_meters

for step in range(num_steps):
    current_time = step * dt
    time_points.append(current_time)

    # Generate new particles at the left edge
    if step % generation_interval_left == 0:
        new_positions_left = np.array([0.0] * particles_per_edge_left)
        new_particle_ids_left = np.arange(next_particle_id, next_particle_id + particles_per_edge_left)
        next_particle_id += particles_per_edge_left
        # Update positions and particle IDs
        positions = np.concatenate((positions, new_positions_left))
        particle_ids = np.concatenate((particle_ids, new_particle_ids_left))

    # Generate new particles at the right edge
    if step % generation_interval_right == 0:
        new_positions_right = np.array([L_meters] * particles_per_edge_right)
        new_particle_ids_right = np.arange(next_particle_id, next_particle_id + particles_per_edge_right)
        next_particle_id += particles_per_edge_right
        # Update positions and particle IDs
        positions = np.concatenate((positions, new_positions_right))
        particle_ids = np.concatenate((particle_ids, new_particle_ids_right))

    # Record current positions
    positions_over_time.append(positions.copy())
```

```
```
AI-driven random walk simulations of viscophoresis and visco-diffusiophoretic particle trapping

```python
        # Record positions of tracked particles
        for pid in trajectories.keys():
            if pid in particle_ids:
                idx = np.where(particle_ids == pid)[0][0]
                trajectories[pid].append(positions[idx] * 1e6)  # Convert to micrometers
            else:
                trajectories[pid].append(np.nan)  # Particle has vanished

        # Calculate local diffusivity D(x)
        D_x = D_start + (D_end - D_start) * (positions / L_meters)

        # Calculate local diffusiophoretic drift v(x)
        v_x = v_start + (v_end - v_start) * (positions / L_meters)

        # Include drift term due to position-dependent diffusivity
        drift = alpha * D_gradient * dt

        # Generate stochastic term
        stochastic_term = np.random.normal(0, np.sqrt(2 * D_x * dt))

        # Update positions with drift, stochastic term, and diffusiophoretic drift
        positions += drift + stochastic_term + v_x * dt

        # Remove particles that have moved out of the range [0, L_meters]
        in_bounds = (positions >= 0) & (positions <= L_meters)
        positions = positions[in_bounds]
        particle_ids = particle_ids[in_bounds]

# Convert time_points to numpy array for plotting
time_points = np.array(time_points)
```